# Strong Interaction between Surface Plasmons and Chiral Molecules

Yangzhe Guo[1], Guodong Zhu[1] and Yurui Fang[1,*]

[1.]*Key Laboratory of Materials Modification by Laser, Electron, and Ion Beams (Ministry of Education); School of Physics, Dalian University of Technology, Dalian 116024, P.R. China.*

*Corresponding author: [Yurui Fang (yrfang@dlut.edu.cn)](mailto:yrfang@dlut.edu.cn)*

**Abstract**

In plasmonic chirality, the phenomenon of circular dichroism for achiral nanoparitcles caused by Coulomb interaction between metal nanoparticles (NPs) and chiral molecules have been studied. At the same time, under the resonance condition, the dye molecules and metal NPs will produce huge Rabi splitting due to strong coupling. If the chiral molecules are at the resonance of the plasmon, what will happen for the strong interaction between the plasmon and molecules with chirality introduced? In this paper, we investigate a spherical core-shell model and analyze its spectral phenomena under the excitation of circularly polarized light (CPL). Based on Coulomb interaction between NPs and chiral molecules, we will show how the various factors affect the strong coupling. We have obtained three mechanisms for the interaction between plasmons and chiral molecules: strong coupling (Rabi splitting up to 243mev), enhanced absorption and induced transparency. The interaction between CPL and chiral molecules with the opposite chirality to CPL is stronger than that of the same chirality, and the line width of the two peaks is closer than that of the same chirality, which shows that for the Rabi splitting with chirality, there are deeper mechanisms for the interaction. This result will be helpful for further research on the interaction between plasmon and molecules with chirality.

**Introduction**

Chirality is a general phenomenon in nature and life. The main characteristic of chirality is that the structure cannot coincide with its mirror by simple rotation or translation [1]. Protein and nucleotide have chirality which is the dominant component of life. Optical activity (OA) is a very important property of the different response of the chiral object under the excitation of left circularly polarized light (LCP) and right circularly polarized light (RCP)[2-4]. Circular dichroism (CD) and optical rotatory dispersion (ORD) are powerful spectral tools for recording, analyzing and manipulating the molecular chirality in the fields of physics, biology, medicine and chemistry[5-9]. Generally speaking, the OA of nature chiral molecules is very weak so that it's difficult to interact strongly with light[10]. So it needs high concentration or large analyze volumes to study these properties, while plasmonic metamaterial can enhance the interaction between light and matter[11]. The collective oscillation of free electrons at the metal and dielectric interface[12, 13] shows the ability of light confinement and field enhancement[14, 15]. It can enhance the interaction between light and matter, resulting in the enhancement of chiral response[16]. Utilizing the surface enhanced feature of noble metal nanostructure, scientists have been able to achieve stronger light-matter interaction. Nowadays, many nanoscale chiral structures have been assembled, and their CD and ORD are obvious, such as metal helix[17-19], two overlapping metal strips[20, 21], the nanostructure of gammadion shapes[22, 23], DNA-based assembled gold particles[6, 24], oligomers of nanodisks[25, 26] and so on[27, 28].

Efficient light-matter interaction is also a key method to achieve the verification of basic physical image, ultra-sensitive sensing, and manipulation of chemical molecular reaction and so on. Realization of strongly light-matter interaction in nano scale or even in the level of single molecule can support an efficient method not only for the breakthrough in optical catalysis and research on micro optical mechanism but also for the application of enantiomers separation in life sciences[29, 30]. Especially in the last decade, some considerable achievements have been obtained that benefiting from the extremely field enhanced feature of surface plasmon in strongly interaction of light field with molecule exciton[31-44]. Due to the extremely field enhancement feature, over the past dozen years, meta-molecule, meta-surface and meta-material with chiral nano structure for enhancing light-matter interaction in sub-wavelength scale has been processed[29, 30]. It has been shown that the metal core-chiral molecule shell plasmon resonance can significantly enhance CD[45]. This enhancement mechanism can be explained by the Coulomb dipole-dipole interactions between a chiral molecule and a nanoparticle (NP). Alexander O. Govorov has theoretically explained this mechanism and obtained the corresponding relation of CD[46, 47]. When light interacts with matter, there will be strong or weak coupling between the plasmon and exciton[48-50], which reflects the phenomena of Rabi splitting[51-54], Fano resonance[55-58] and so on[59].

Considering the importance of chirality and strong coupling in fundamental physics study, one should expect extra essential mechanism during the interaction of a coupled chiral molecule and plasmonic structure system. In this paper the interaction between light and matter is studied from their parameter-dependent spectral phenomena with a core-shell system composing of plasmonic core and chiral molecules shell. The results show that geometric factors affect core-shell coupling; there are three mechanisms for the coupling; in the strong coupling region, there is slightly difference for the chiral core-shell system under LCP and RCP excitation which related to the oscillator strength and chirality. The interaction between the chiral molecule and the metal sphere is stronger for the L- molecule under LCP excitation, which shows larger Rabi splitting peaks and the life time of the higher energy peak becomes longer meantime the life time of the lower energy peak becomes shorter. This result will be helpful for further research on the interaction between plasmon and chiral molecules in plasmon hybridization and energy level survival life.

**Theoretical Considerations**

In this paper, we use a spherical core-shell structure with a core radius of $r$, a shell radius of $h$ and the refractive index of the surrounding medium of 1.3. The core is Ag described by Drude model $\varepsilon_c(\omega) = \varepsilon_P - \omega_P^2/(\omega(\omega + i\gamma))$. The parameters are selected by fitting Johnson and Christy's data[60] in the $400 - 600 \, nm$ range which are $\varepsilon_P = 3.7$, $\hbar\omega_p = 8.55 \, eV$, $\hbar\gamma = 65 \, meV$.

The molecules we choose are chiral, which include right handed and left handed. The permittivities are $\varepsilon_R = (\xi_c + \sqrt{\varepsilon_c + \xi_c^2})^2$ and $\varepsilon_L = (-\xi_c + \sqrt{\varepsilon_c + \xi_c^2})^2$ representing right handed structured (D) molecule and left handed structured (L) molecule, respectively. $\varepsilon_c$ and $\xi_c$ are parameters of dielectric and chiral properties[47].

$$\varepsilon_c = \varepsilon_\infty - \gamma_c \left( \frac{f}{\hbar\omega - \hbar\omega_{ex} + i\gamma_{ex}} - \frac{f}{\hbar\omega + \hbar\omega_{ex} + i\gamma_{ex}} \right) \tag{1}$$

$$\xi_c = \beta_c \left( \frac{f}{\hbar\omega + \hbar\omega_{ex} + i\gamma_{ex}} + \frac{f}{\hbar\omega - \hbar\omega_{ex} + i\gamma_{ex}} \right) \tag{2}$$

We define $\varepsilon_\infty$ is the large-frequency permittivity and set as 1.7, $f$ is a dimensionless oscillator strength. $\gamma_c = 4\pi n_0 \frac{\mu_{12}^2}{3}$ is coefficient that determine the magnitude of absorptive properties. $\omega_{ex}$ and $\gamma_{ex}$ are resonance position and resonance width. $\mu_{12} = 0.1225 \, e\text{Å}$ is electric dipole moment where $n_0$ is the density of molecular dipoles and $e$ is elementary charge.

$\beta_c = 4\pi n_0 i \frac{(\mu_{12} \cdot m_{12})}{3}$ is the coefficient that determine the magnitude of chiral properties where $m_{12} = 9.125 \times 10^{-4} \, e\text{Å}$ is magnetic dipole moment represented with electric dipole moment for simplifying the calculations.

We use the commercial software (COMSOL Multiphysics 5.4) based on finite element method (FEM) to calculate all physical conditions in the following part of this paper. The spherical core-shell structure is in a $900 \, nm$ thick sphere (including core-shell) homogeneous medium with $n = 1.3$ and the outermost sphere layer is a perfect matched layer (PML) with a thickness of $300\ nm$ which is a complete absorber of the scatter field. All the above structures are constructed with mesh no more than $\lambda/6$ (tetrahedral mesh for domain and triangular for face.) and the maximum mesh size of the core-shell structure is less than $5\ nm$. The scatting cross section is defined as the surface integral of the dot product of the normal vector pointing outwards from the core-shell and the scattered intensity (Poynting) vector on a closed sphere surface with a radius of $300\ nm$ divided by the incident intensity. The absorption cross section is the volume integral of the power loss density in the particle divided by the incident intensity. And the extinction cross section is the sum of the above.

**Coupling Strength**

Figure 1a, b shows extinction cross section spectra of the core-shell structure under the excitation of LCP light (The RCP excitation is also calculated and almost the same as that of LCP excitation). The radius of the core is 15 nm to 35 nm. The thickness of shell is 2 nm. Dimensionless oscillator strength and molecule resonance width are $f = 1$ and $\hbar\gamma_{ex} = 50\ meV$, respectively. In the calculation, resonance position of the shell is consistent with variable that of core (variable radius). The results show that the two peak spectrum redshifts with the increase of radius. The decrease between two peaks indicates the strong coupling between the plasmon and the exciton. The coupling strength determines the regime. By changing the radius of the core from 15 nm to 35 nm, the shell to core volume ratio changed from 0.46 to 0.18. Thus, the ratio of the oscillation strength changed so that the interaction will decrease as radius increases. For the core-shell structure with small radius, there is a large peak splitting due to the strong interaction of the plasmon exciton (see the normalized extinction spectrum).

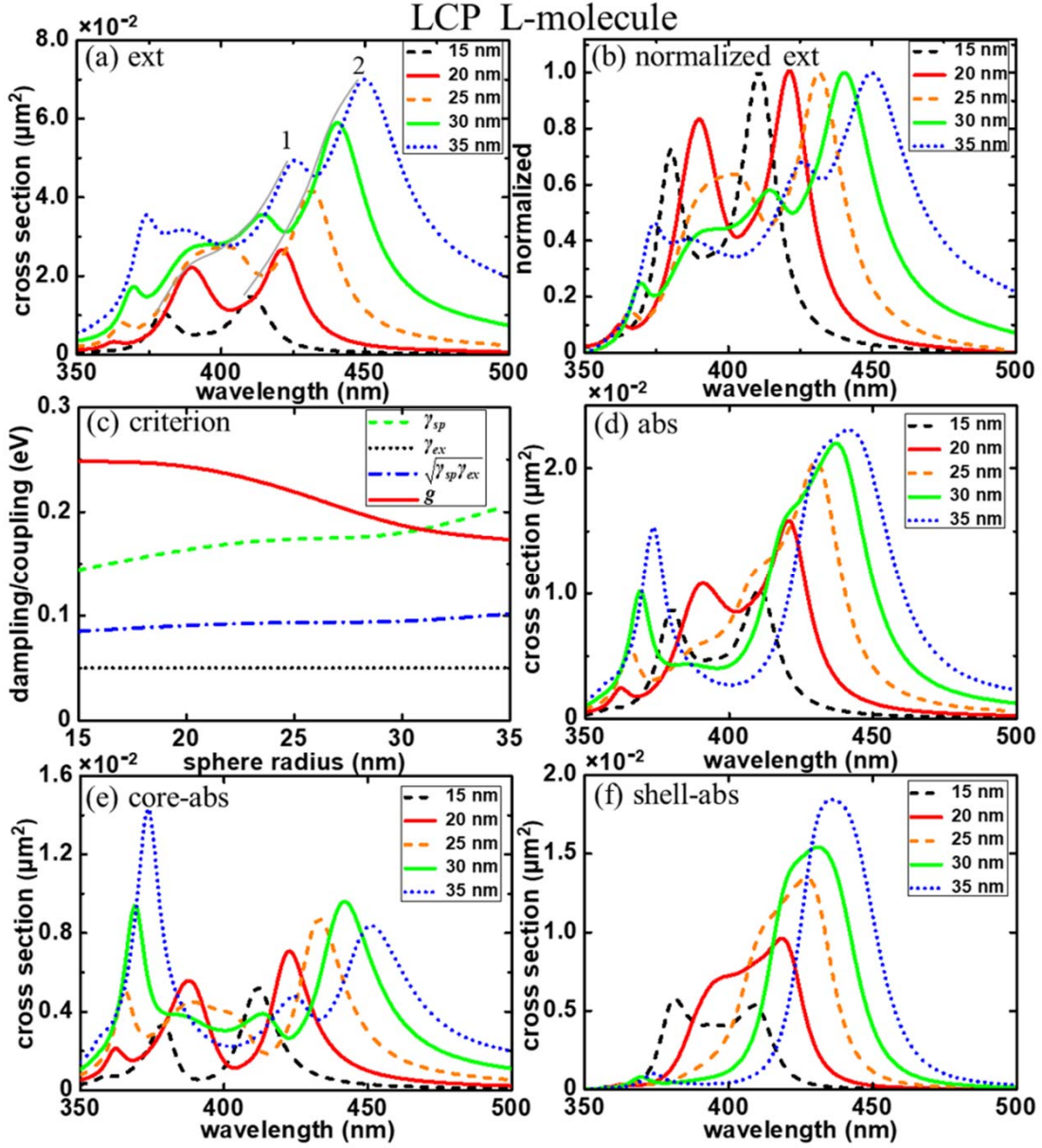

Figure 1. (a) Extinction cross section spectra of the structure of silver core with radius from $15\ nm$ to $35\ nm$ and left-handed molecular shell with thickness of $2\ nm$ which is excited by LCP. (b) Corresponding normalized extinction spectra. (c) Line width and coupling strength diagram of plasmon and exciton. According to criterion $g > (\gamma_{ex}, \gamma_{sp})$, it shows that it is strong coupling when the radius is less than $30\ nm$. (d-f) Corresponding total, core and shell absorption cross section spectra with the same structure as (a). The peak around 370 nm for the 35 nm particle is higher order resonant peak.

A coupled harmonic oscillator model is used to calculate coupling strength $g = \sqrt{\Omega_R^2 + \frac{(\gamma_{sp}-\gamma_{ex})^2}{4}}$ and the line width of plasmon resonance and exciton (see in Figure 1c)[61]. The relation for strong coupling of $g, \gamma_{ex}$ and $\gamma_{sp}$ has been given as $g > (\gamma_{ex}, \gamma_{sp})$ or $g > (\gamma_{ex}\gamma_{sp})^{\frac{1}{2}}$ [54]. As the radius increases, exciton dephasing rate smaller than $g$, while the plasmon line width increases gradually. When $r > 30\ nm$, $\gamma_{sp}$ will be larger than $g$, and the system is in a weak coupling mechanism. In contrast, when $r < 30\ nm$, the interaction of the system is very strong. In Figure 1d-f, by comparing the three absorption spectrum, we find out that the inter band absorption in the high energy region around $360 - 370\ nm$ increases with the increase of radius mainly contributes from the core and this peak is a higher order resonant peak . As the radius increases to $30\ nm$, the absorption spectrum of the core decreases from deep to shallow, from wide to narrow, while the two peaks in the absorption spectrum of the shell gradually become one peak, and the absorption of the shell becomes larger and occupies the dominant position in the total absorption.

From the spectrum above, one can find two typical coupled resonant cases from $r = 15nm$ in Figure 2a-b to $r = 35nm$ in Figure 2c-d that represent the strong coupling and enhanced absorption, respectively. Figure 2a, c show the scattering and extinction spectra at $15\ nm$ and $35\ nm$ respectively (solid line represents the core-shell structure and short dash line represents the bare core structure). And the corresponding absorption spectra are also shown in Figure 2b, d (solid line represents the core-shell structure and short dash line represents the bare core structure).

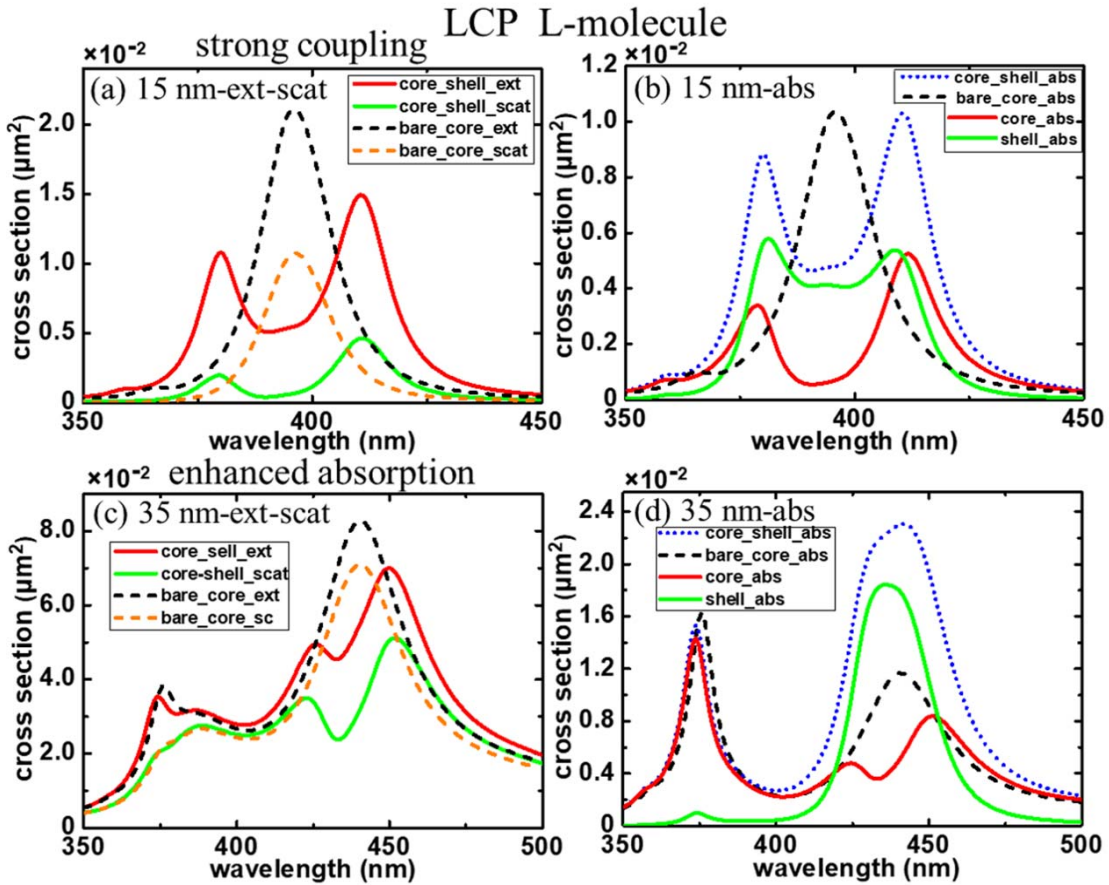

Figure 2. Optical spectra of the core-shell sphere structure with the radius of silver core (a, b) $r = 15\ nm$ (plasmon resonance wavelength is about $395\ nm$), (c, d) $r = 35\ nm$ (plasmon resonance wavelength is about $440\ nm$) and a left-hand molecule layer (molecule resonance is the same as the plasmon) whose thickness fixed at $2\ nm$. (a, c) Scattering and extinction spectra of core-shell structure (solid line) and the bare core structure (short dash line). (b, d)

Absorption spectra of the above two structures (solid line represents the core-shell structure and short dash line represents the bare core structure).

Due to the strong coupling mechanism, the absorption, scattering and extinction spectra of core-shell structure with $r = 15\ nm$ have a large splitting of $30.63\ nm$ ($243\ \text{meV}$) which is an obvious characteristic of the strongly interaction between plasmon and exciton. Because of the quite suppression of scattering at resonance energy ($395\ nm$), the absorption spectrum accounts for a large proportion of the extinction. The absorption of individual core and shell are shown in Figure 2b. The part of core in the core-shell structure also shows a huge splitting, but the part of shell is different from core. Although the absorption of the shell also has two splitting peaks, there is no dip like the core at the resonance. On the contrary, the absorption is high and shows a horizontal spectrum between the two peaks.

Figure 2c, d shows spectra of core-shell structure with core radius of $35\ nm$. It indicates that the value of scattering at the dip ($440\ nm$) is about one third of the peak of bare core and the extinction at the dip is exceeds half of the peak of bare core. This phenomenon of reduced splitting compared to strong coupling points to the weak coupling mechanism. Meanwhile, instead of splitting at the plasmon resonance strictly as we expected, the position of dip in all three spectra is blue-shifted compared with resonance position. Which leads to the difference of the line width of the two peaks and affects the lifetime of the two splitting energy level. When observing the absorption in the core-shell structure, the extinction is no longer dominated by the absorption like strong coupling. The splitting only occurred in the core absorption, and neither the shell nor the total absorption. It is precisely because the absorption of the shell at resonance is increased (1.6-fold bare core) without splitting, the splitting of core absorption is also reduced, resulting in an increase in total absorption (2-fold bare core) without splitting. Obviously, due to the change of geometry in core-shell structure, the weaker plasmon-exciton interaction leads to an enhanced absorption mechanism.

**Interaction with chirality**

Although chiral molecule was use in above analysis, only the coupling regions were focused on. The chirality of the molecule shell should have influence for the coupling effect (Figure S1). From the permittivity expression of the chiral molecule one can see that there are several parameters significantly affect the chirality which surely related to the coupling strength. The effect of the oscillation strength $f$ is investigated. The oscillation strength $f$ represents the amplitude or the equivalent number of oscillators for each molecule. Figure 3 shows the spectrum of core-shell structure ($r = 20\ nm$, $\hbar\gamma_{ex} = 50\ meV$) as the oscillator strength changes from $0.25$ to $3$. When the oscillator strength increases, the two peak splits more and more, which further affects the damping of plasmon. When $f > 0.5$, it is a strong coupling mechanism. When $f > 2$, there is a third peak between the two peaks. This similar phenomenon occurs when the dye molecules are excited by linearly polarized light (LPL), and we also find this phenomenon when the chiral molecules are excited by CPL. At the same time, when different CPL is incident, the peak positions are different. The three peaks of LCP are redshift than that of RCP, which may be caused by the different interaction between left-handed molecules and different CPL. Figure 3c shows the difference in Figure 3b for L-molecule core-shell under LCP and RCP excitation. Both of the difference of the two split peaks decrease when $f$ increases from 0.25 to 1. When $f$ increases from 1 to 3, the left peak difference increases but the right peak difference almost keeps the same. When the middle peak shows up, the difference also increases as $f$ increasing.

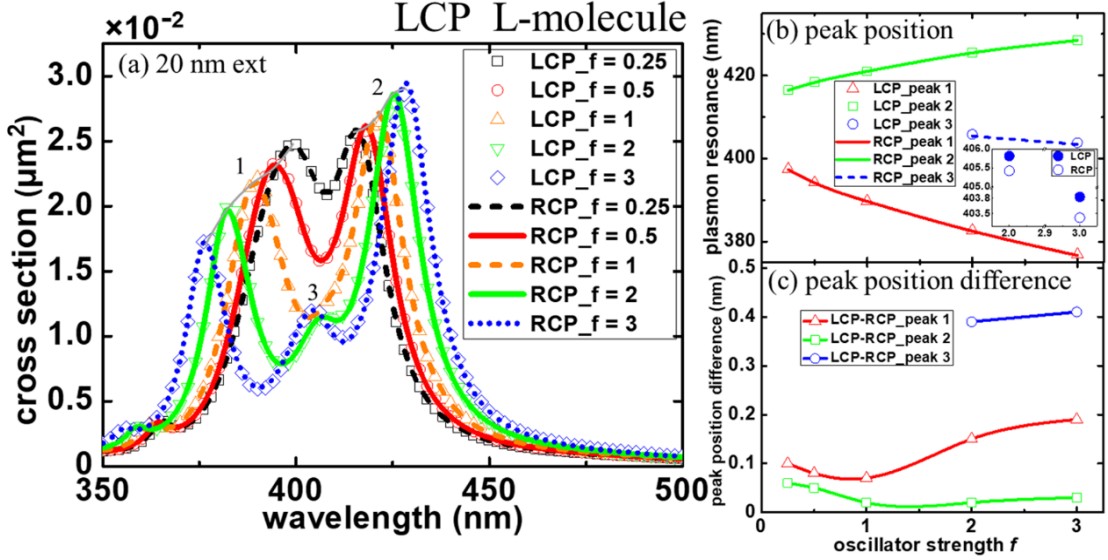

Figure 3. (a) Extinction cross section spectra of silver core with radius of $20\ nm$ (plasmon resonance wavelength is about $410\ nm$) and $2\ nm$ thick left-handed molecular shell with different oscillator strength $f$ from 0.25 to 3 excited by LCP and RCP. (b) Peak position chart corresponding to extinction spectra (a). The illustration is a magnification of peak 3. (c) Peak position difference between excited by LCP and RCP.

In the region of oscillator strength $f$ between 0.25-1, when $f$ is reduced, the splitting of extinction spectra is reduced, indicating that the coupling strength is reduced. And the absorption spectra of the shell become almost one peak when $f = 0.5$ (as shown in Figure 4b). It is obvious that the splitting of scattering and extinction at resonance is between strong coupling and enhanced absorption mechanism. The dip of scattering and extinction is higher than that of strong coupling, which shows that the interaction is weak. While the total absorption is equivalent to bare core and lower than the enhanced absorption. Excluding the absorption of the shell which dominates the total absorption, the core absorption has a deep dip at resonance. The small core absorption and the total absorption equivalent to the bare core in this core-shell structure can be defined as the induced transparency mechanism. When the oscillator strength $f = 3$ with a radius of $20\ nm$. We have obtained absorption, scattering, and extinction optical cross sections (Figure 4c). Three peaks structure can be seen in all three spectra, which shows that the structure is strongly coupled. Because of the strong coupling and large oscillator strength, the inner part of the molecules interacts strongly with the core and the outer part has very weak interaction with the core. So the third peak is just the resonance of the molecule.

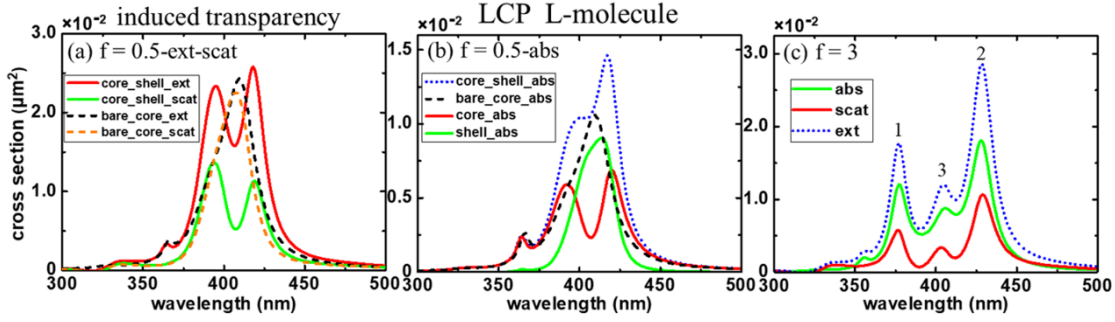

Figure 4. Optical cross section of nanoparticles with silver core having $R = 20\ nm$ (plasmon resonance wavelength is about $410\ nm$) and a left-hand chiral molecule layer with $2\ nm$ thickness ($f = 0.5$, $\hbar\gamma_{ex} = 50\ meV$). (a) Comparison of extinction and scattering of core-shell (solid line) and bare core (short dashed line) structures. (b) Comparison of absorption of core, shell (solid line), core-shell (short dotted line) and bare core (short dashed line) structures. The absorption of the core decreases at the plasmon resonance, and the shell absorption is dominating which is defined as the induced transparency mechanism. (c) Optical cross section of nanoparticles with silver core having $R = 20\ nm$ and a left-hand chiral molecule layer with $2\ nm$ thickness ($f = 3$, $\hbar\gamma_{ex} = 50\ meV$). The absorption (solid green line), scattering (solid red line) and extinction (short dotted blue line) spectra are depicted respectively and all have three peaks.

By keeping the plasmon resonance at a constant wavelength and changing the resonance wavelength of the exciton, the characteristic anti-crossing spectra indicating the coupling state can be obtained. As shown in Figure 5a, the fixed core radius is $15\ nm$ ($f = 1$), the constant plasmon frequency is fixed, and the resonance wavelength of the exciton is varying from $375\ nm$ to $425\ nm$. The core-shell structure extinction spectra for LCP excitation is obtained (The RCP excitation is also calculated and almost the same as that of LCP excitation). We collect the fitted linewidth data (Figure 5b). The two peak linewidth values of RCP are closer than that of LCP, that is, the left peak of RCP is wider and the right peak is narrower. It indicates that when the resonance the core and shell overlap, the life time of the both peak modes becomes longer and then increase. But the first peak mode has narrower width indicating longer life time than the second peak mode.

The obvious anti-crossing of the extinction spectrum is observed. In this case, the peak position of extinction spectra forms Rabi splitting (Figure 5c-d). According to Lorentz line fitting results (Figure 5c-d), it is found that at the initial $375\ nm$, the left peak of RCP is blue shifted compared with LCP (Figure 6a-b), while the right peak is red shifted compared with LCP, which is equivalent to that the left-handed molecules respond more strongly to RCP, and the interaction is stronger. With the increase of resonance wavelength, the right peak of RCP is blue shifted gradually to the left side of LCP, and the spectra at this time is equivalent to that RCP shifts to the left compared with LCP.

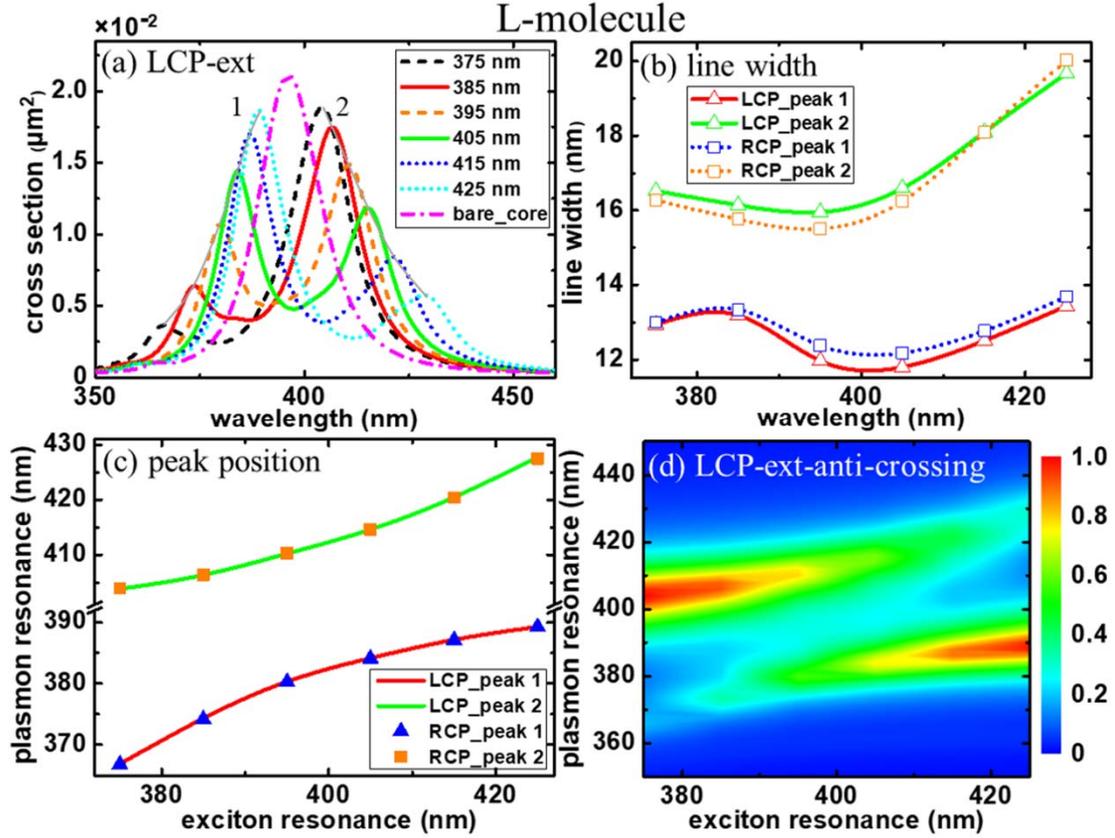

Figure 5. (a) Extinction cross section spectra of silver core with radius $15\ nm$ (plasmon resonance wavelength is about $395\ nm$) and $2\ nm$ thick left-handed molecular shell with different molecular resonance wavelength from $375\ nm$ to $425\ nm$ excited by LCP. (b) Linewidth chart and (c) peak position chart which collected from extinction cross section spectra excited by LCP and RCP. (d) Normalized anti-crossing spectra corresponding to (a).

If investigating the difference of the split peaks for LCP and RCP light, it can be found that for the left peak, the difference becomes smaller as molecule resonance sweeping from high energy to low energy (Figure 6a). However, for the right peak, the peaks for LCP and RCP cross each other (Figure 6b). And the left peak difference changing is much smaller than the right peak difference (Figure 6c). For the peak width, one can find that for both peak 1 and peak 2, the difference appears at $395\ nm$ where the resonances of the molecule and plasmonic particle overlap (Figure 6d, red and black lines). It indicates the life time difference for the same peak under LCP and RCP are maximum at the resonance overlapping position. And from the orange and green lines, one can find that the peak 1 always has narrower line width because the higher energy peak is a sub-radiant mode (Figure S3). From Figure 6e one can also find out that the total energy dissipation increases if the resonances of the molecules and plasmon are detuned. The CD spectra in Figure 6f shows that as the resonances detuning increases, the CD becomes smaller.

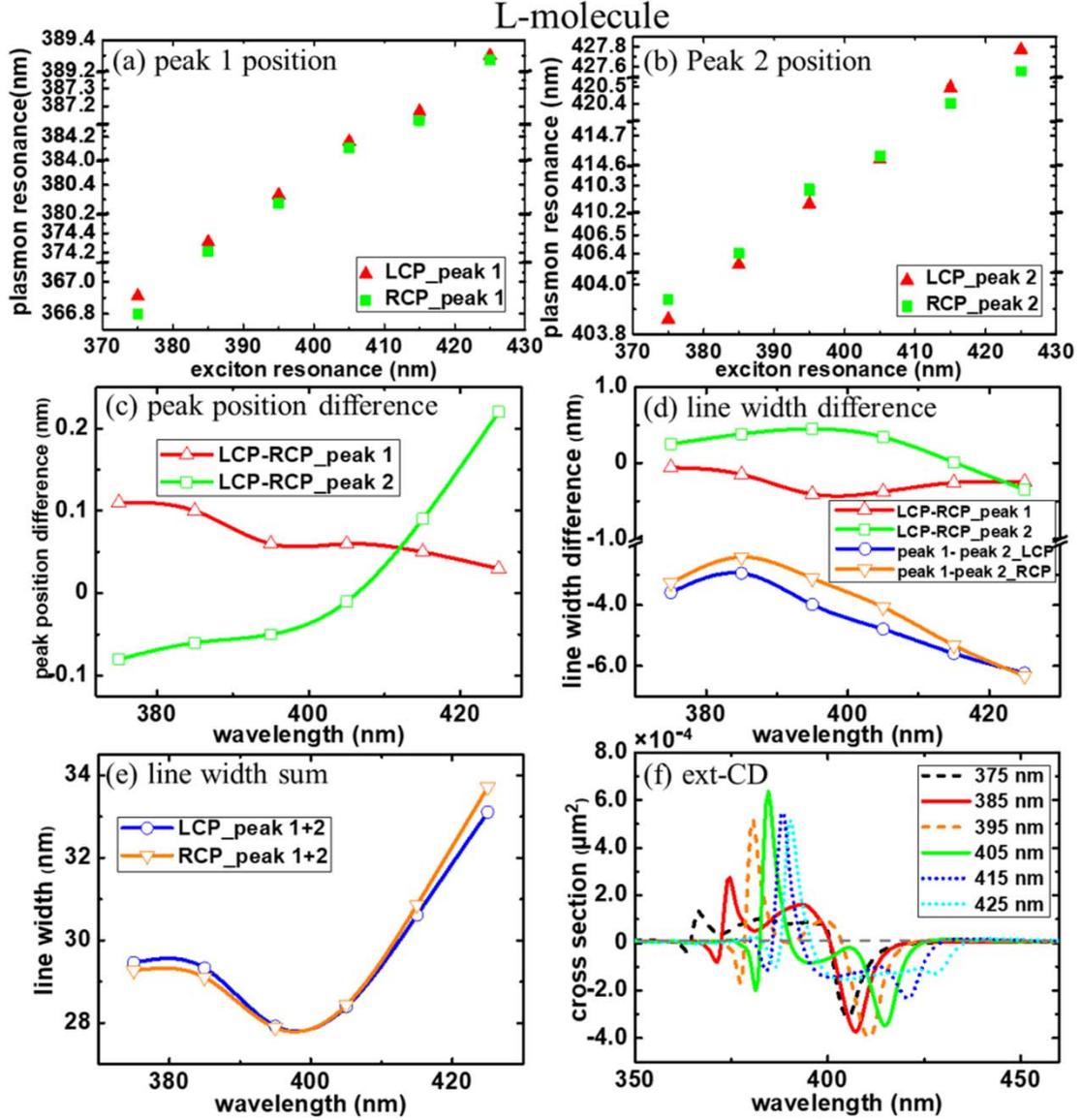

Figure 6. (a) Peak 1 and (b) peak 2 position of silver core with radius $15\ nm$ (plasmon resonance wavelength is about $395\ nm$) and $2\ nm$ thick left-handed molecular shell with different molecular resonance wavelength from $375\ nm$ to $425\ nm$ excited by LCP and RCP. (c) Position of peak 1 difference and peak 2 difference under LCP and RCP excitation, respectively. (d) Four difference combinations of line width. (e) Line width sum of peak 1 and peak 2 under LCP and RCP excitation. (f) The corresponding extinction spectra of CD.

In order to further investigate the chiral plasmon-exciton interaction, the electric dipole moment is enlarged into 4 time of previous value as $\mu'_{12} = 0.49\ e\text{Å}$ and other parameters remain unchanged. So the permittivity of the molecule increases but relatively the chirality decreases because the magnetic dipole moment is unchanged. The molecule resonance wavelength varies from $375$ to $415\ nm$. The extinction spectra of core-shell structure are shown in Figure 7. The splitting gap is much larger than previous results in Figure 5a (as shown in Figure 7a) and obvious anti-crossing of the extinction spectrum is observed (Figure 7c and Figure S4c). According to the Lorentz line fitting results, it is found that the linewidth values of the two peaks are closer when RCP is incident than that of LCP, that is, the left peak of RCP is wider compared with LCP and the right peak is narrower (Figure 7b).This is consistent with the result of $\mu_{12} = 0.1225 \ e\text{Å}$. Similarly, the achiral molecules ($hm = 0$) excited by LCP are also calculated. The result shows that the linewidth (Figure 7b) and peak position (Figure S4a-b) are between the chiral molecules excited by LCP and RCP. The strong coupling makes the Rabi splitting covering larger wavelength range and almost the same splitting peak gap. The strong coupling also makes the CD varying smaller in the Rabi splitting range (Figure 7d).

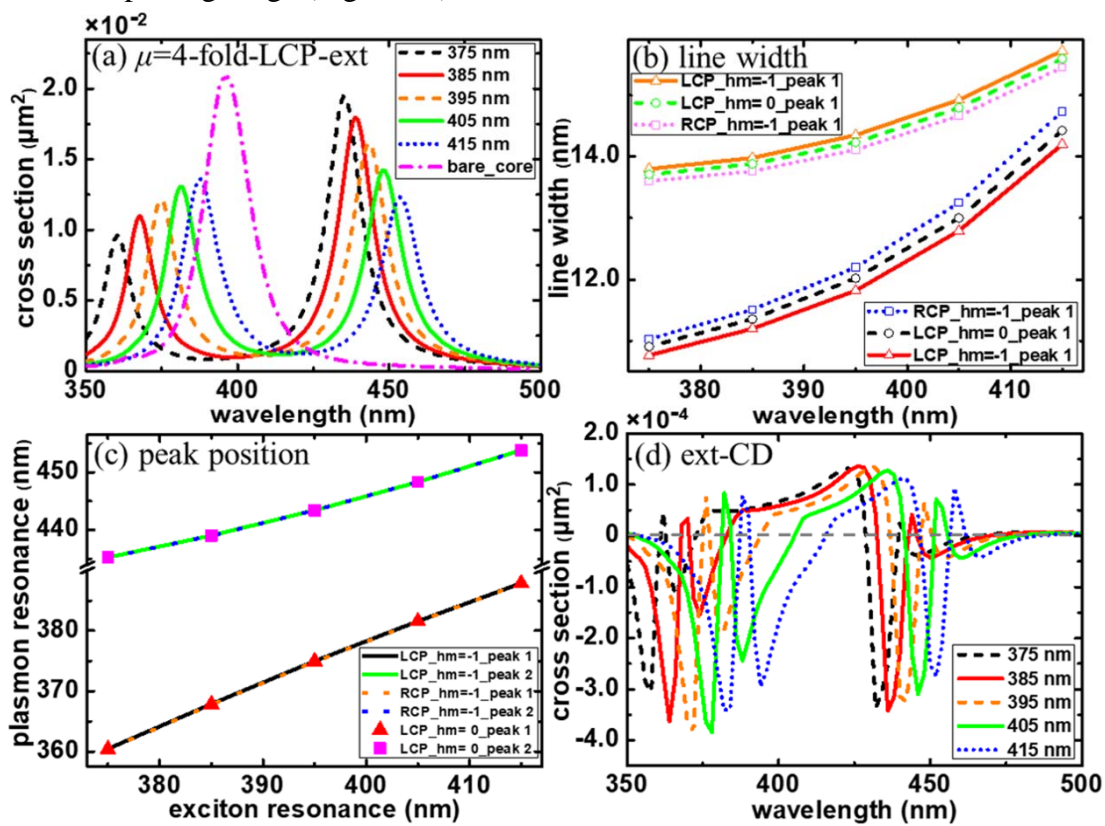

Figure 7. (a) Extinction cross section spectra of silver core with radius $15 \ nm$ (plasmon resonance wavelength is about $395 \ nm$) and $2 \ nm$ thick left-handed molecular shell with different molecular resonance wavelength from $375 \ nm$ to $415 \ nm$ excited by LCP but change the electric dipole moment to $\mu_{12}^{'} = 0.49 \ e\text{Å}$. (b) Linewidth chart, (c) peak position chart and (d) CD corresponding to extinction spectra.

**Conclusion and Discussion**

In conclusion, we use numerical simulation to study the influence of geometric factors on the coupling of plasmon and chiral molecules with spherical core-shell structure. Three interaction mechanisms are found by coupling criterion. Then, a series of phenomena are found and explained by changing the oscillator strength, resonance frequency and chirality of chiral molecules. Finally, the most important thing is to find that different chirality has a great influence on the resonant linewidth of the split two peaks, which leads to the significant difference of the two-level lifetime. We have obtained three mechanisms for the interaction between plasmons and chiral molecules: strong coupling (Rabi splitting up to 243mev), enhanced absorption and induced transparency. The interaction and splitting become stronger with the decrease of radius, the increase of oscillator strength and electric dipole moment. The interaction between CPL and chiral molecules with the opposite chirality to CPL is stronger than that of the same chirality, and the line width of the two peaks is closer than that of the same chirality. The peak position and line width of the achiral molecule are between the above two situations. CD increases as it approaches plasmon resonance. Superficially, the strong coupling difference of the chiral molecule and plasmonic sphere for LCP and RCP is due to the slightly permittivity difference of the molecule. However, there should be more profound mechanism in such system. A more precise picture should be draw with quantum optics theory. It is expected that the Rabi frequency $g = \sqrt{\Omega_R^2 - \Delta^2}$ should related with more parameters contain chirality properties like $g = \sqrt{\Omega_R^2 + \frac{(\gamma_{sp} - \gamma_{ex})^2}{4} - \Delta^2 \pm D^2}$, where $\Delta = \omega_{sp} - \omega_{ex}$, $D = G'' * C/\hbar$. The parameter $G''$ is the mixed electric-magnetic polarizability [62] of the molecule which reflects the optical activity response of the chiral molecule. $C$ is the chiral near field [62] yielded by the nanoparticle under CPL excitation which responses the interaction of the chiral part of the molecule.

## Acknowledgement

## Funding
National Natural Science Foundation of China (NSFC) (11704058); Fundamental Research Funds for the Central Universities (DUT19RC(3)007).

## Conflicts of interest
The authors declare no competing financial interest.

# Supporting information for

# Strong Interaction between Surface Plasmons and Chiral Molecules

Yangzhe Guo[1], Guodong Zhu[1] and Yurui Fang[1,*]

[1.]*Key Laboratory of Materials Modification by Laser, Electron, and Ion Beams (Ministry of Education); School of Physics, Dalian University of Technology, Dalian 116024, P.R. China.*

*Corresponding author: Yurui Fang (yrfang@dlut.edu.cn)

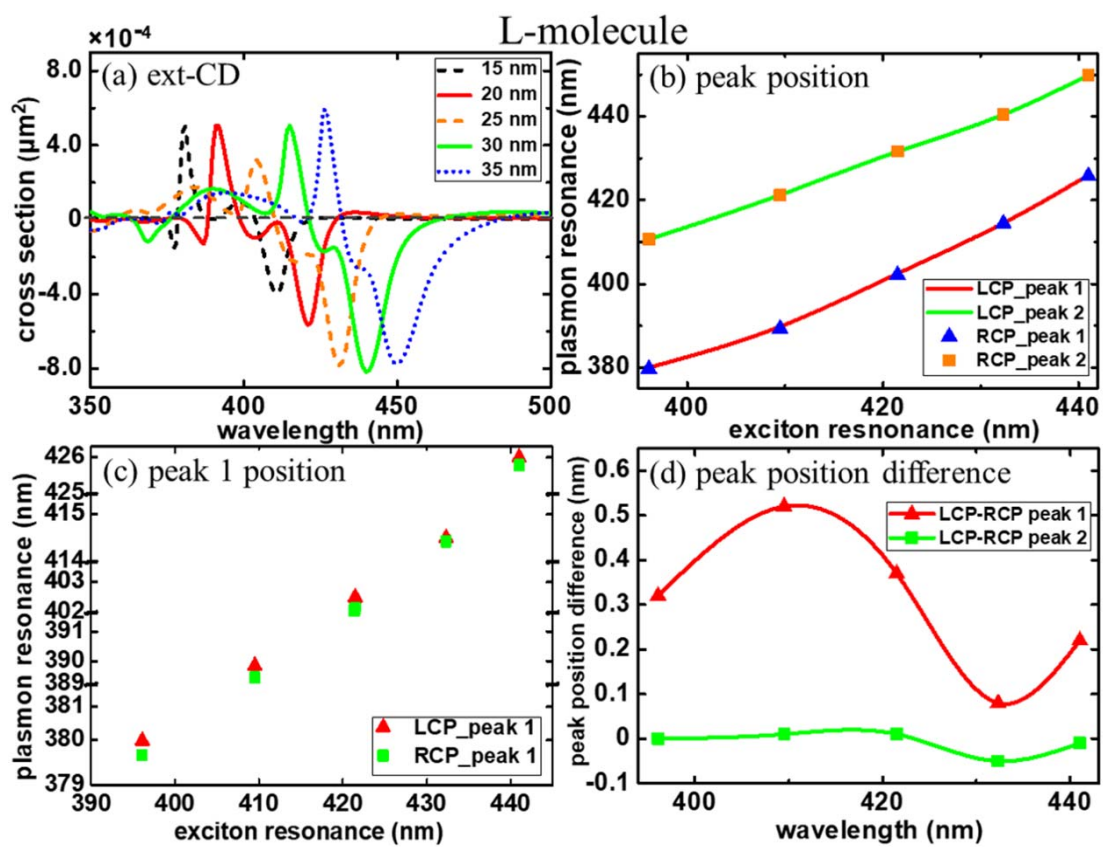

Figure S1. (a) CD of extinction cross section spectra of the structure of silver core with radius from $15\ nm$ to $35\ nm$ and left-handed molecular shell with thickness of $2\ nm$ which is excited by LCP and RCP. (b) Corresponding peak position chart. (c) Peak 1 position chart. (d) Peak 1 difference and peak 2 difference under LCP and RCP excitation, respectively.

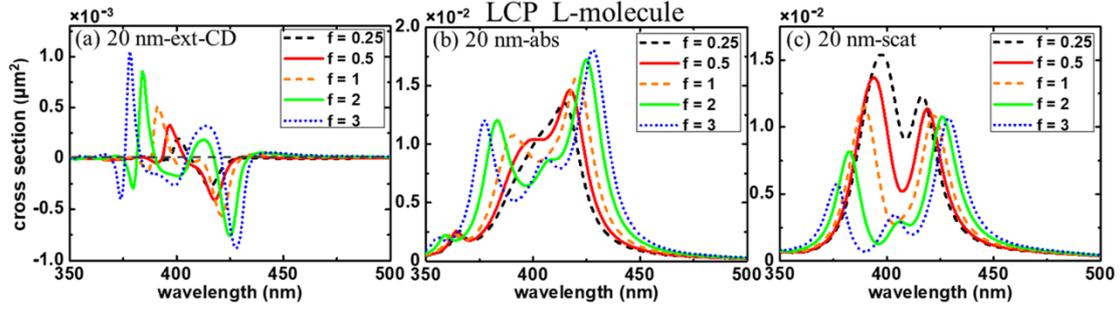

Figure S2. Cross section spectra of silver core with radius of $20\ nm$ (plasmon resonance wavelength is about $410\ nm$) and $2\ nm$ thick left-handed molecular shell with different oscillator strength $f$ from 0.25 to 3 excited by LCP. (a) Corresponding CD of extinction spectra. (b) Absorption spectra and (c) scatting spectra.

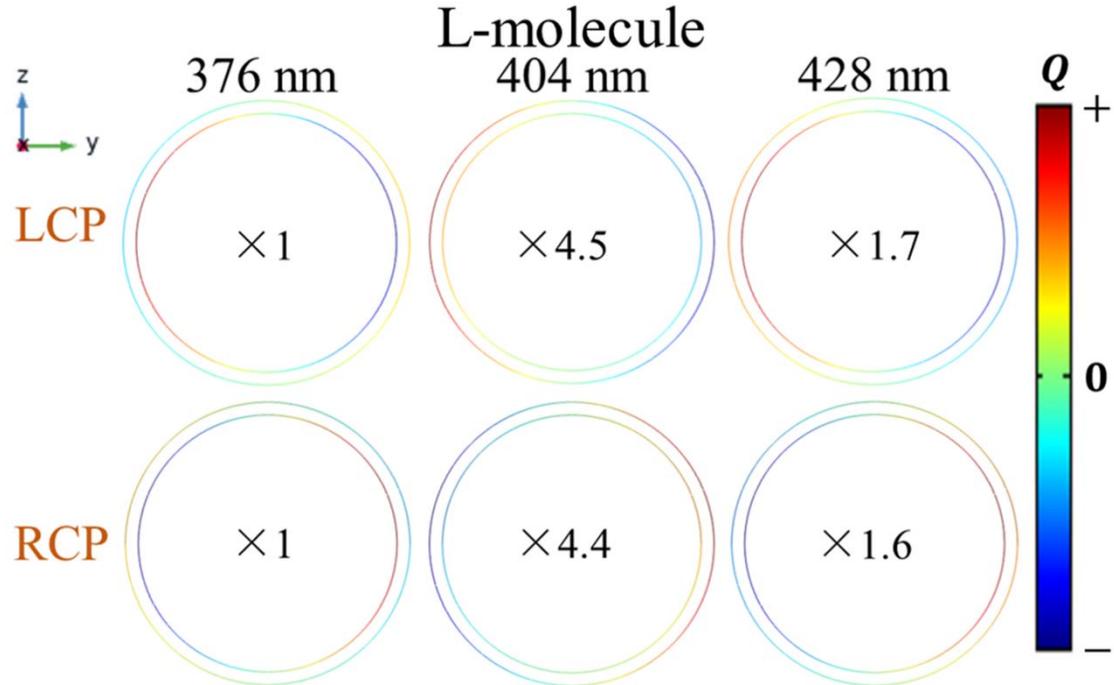

Figure S3. Cross section of surface charge distribution of core-shell nanoparticles under (a) LCP and (b) RCP incidence ($R = 20\ nm$, $f = 3$, $\hbar\gamma_{ex} = 50\ meV$).

In order to clarify the strong plasmon-exciton interaction of the three-peak structure, the charge distribution on the inner (interface between core and shell) and outer (interface between shell and medium) surface of the y-z plane is shown in Figure S3 to facilitate the analysis of the mode. When excited by LCP, the sign of the charge distribution on the inner and outer surface of the peak at $376\ nm$ are opposite, and the charges are mainly concentrated on the inner surface. The distribution of the latter two peaks is different from the former one, and the sign of the inner surface charge distribution is consistent with that on the outer surface. However, the difference in the charge distribution of these two peaks is reflected in the distribution of peak at $404\ nm$ mainly on the outer surface, while the distribution of peak at $428\ nm$ mainly on the inner surface. From the concentrated surface of the charge, the middle peak is obviously different from the other two. Because the charge is concentrated on the outer surface (the interface between molecular shell and medium), the middle peak is more like an exciton resonance peak, which is different from the other two strong coupling peaks (the charge is concentrated on the core-shell interface). The situation of being excited by RCP is the same as that of LCP, except that the opposite charge sign. The more charge concentrated in the latter two peaks (the maximum charge value is larger) also reflects a stronger interaction than LCP to a certain degree.

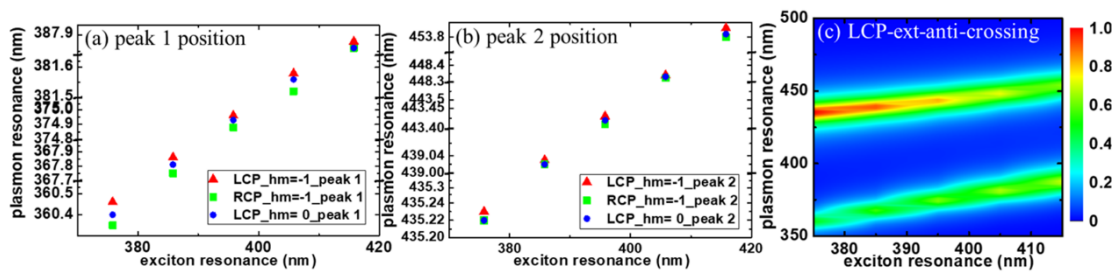

Figure S4 (a) Peak 1 and (b) Peak 2 position of extinction cross section spectra of silver core with radius $15\ nm$ (plasmon resonance wavelength is about $395\ nm$) and $2\ nm$ thick left-handed molecular shell with different molecular resonance wavelength from $375\ nm$ to $415\ nm$ excited by LCP but change the electric dipole moment to $\mu'_{12} = 0.49\ e$Å. (c) Corresponding normalized anti-crossing spectra.